\documentclass[11pt]{amsart}
\usepackage{graphicx}
\usepackage{mathptmx,epsfig,latexsym,epstopdf, subfig}

\begin{document}

\title[Organic Memristor Devices for Logic Elements with Memory]{Organic Memristor Devices \\ for Logic Elements with Memory}

\author[Erokhin, Howard, Adamatzky Int J Bifurcation Chaos 22 (2012) 1250283]
{Victor Erokhin$^1$, Gerard David Howard$^2$, Andrew Adamatzky$^2$}




\maketitle

\centerline{\small $^1$~CNR-IMEM and Department of Physics, University of Parma, Parma, Italy}
\centerline{\small $^2$~University of the West of England, Bristol, UK}

\vspace{0.5cm}

\centerline{\bf Final edited version of the paper is published in}
\centerline{\bf Int J Bifurcation Chaos 22 (2012) 1250283}

\begin{abstract}

Memristors are promising next-generation memory candidates that are nonvolatile, possess low power requirements and are capable of nanoscale fabrication.  In this article we physically realise and describe the use of organic memristors in designing statefull boolean logic gates for the AND OR and NOT operations.  The output of these gates is analog and dependent on the length of time that suitable charge is applied to the inputs, displaying a learning property.   Results may be also interpreted in a traditional binary manner through use of a suitable thresholding function at the output.  The memristive property of the gate allows the for the production of analog outputs that vary based on the charge-dependent nonvolatile state of the memristor.  We provide experimental results of physical fabrication of three types of logic gate.  A simulation of a one-bit full adder comprised of memristive logic gates is also included, displaying varying response to two distinct input patterns.


\end{abstract}


\section{Introduction}

The {\em memristor} (a portmanteau of memory-resistor) was predicted by~\cite{1} and recently fabricated~\cite{2}, which has garnered much attention.   Increased activity in this research field is largely due to the ability of the memristor to open up the possibility of new computing paradigms.  The memristor itself must be a simple element whose conductivity state is determined not by its actual bias conditions but by the ``history'' of its previous functioning. In this respect, a processor realized with such elements will also have memory, mimicking a functioning of nervous system whereby memory is integrated into the processor.  This allows learning, i.e. reconfiguration of the processor hardware according to past experience.  Memristors display interesting computational properties, having previously been used to create oscillators~\cite{oscillators,vic-oscillator} and chaotic circuits, e.g.~\cite{hyperchaos}.

Currently, most reported memristors are based on metal oxide nanostructures, mainly titanium oxide structures~\cite{3}. The main application of these systems is considered to be in the field of non-volatile memory~\cite{4}. However, the realization of logic elements based on memristors has also been reported~\cite{5,6}, including automated search of potential circuit topologies~\cite{Howard}.  From one perspective, it seems counterintuitive to use memristors for a simple reproduction of logic elements that can be easily fabricated with traditional components.  Instead, the unique characteristic property of memristors and memristive systems is the presence of memory. Thus, these elements could establish a new class of electronic logic elements --- logic with memory --- where the output of the element will be not fixed as a binary 0 or 1 signal, but instead will be analog in nature and dependent on the duration and configuration of the input electrodes.

The aim of this work is to realize memristive logic elements. On the one hand, they must perform according to the basic logic function that they represent, such as OR, AND and NOT. On the other hand, the strength of these connections must depend on the duration of appropriate input configurations. In addition, once performed, the configuration must be preserved in the absense of subsequent reinforcement or inhibition.  Such elements will combine properties of the logic with synapse-like memory, imitating to some degree the situation in the brain where the performed decision is determined not only by the current configuration of stimuli, but also by the experience, accumulated during resolving similar problems.

The basic device is an organic memristive system-element, composed of conducting polymer with a solid electrolyte heterojunction~\cite{7}.  Device conductivity is a function of ionic charge which is transferred through the heterojunction~\cite{8,9}. Its application for the realization of adaptive circuits~\cite{10} and systems, imitating synaptic learning~\cite{11}, has been already demonstrated.

In this research we demonstrate physical memristive gates that realise AND OR and NOT operations, and report on their behaviour under testing in the lab.  We further include a simulated one-bit full-adder whose performance is based on the previous experiments and show the circuit learning over time to respond correctly to two separate input configurations.

\section{Materials and Methods}

\begin{figure*}[ht!]
\begin{center}

\subfloat[]{ \psfig{file=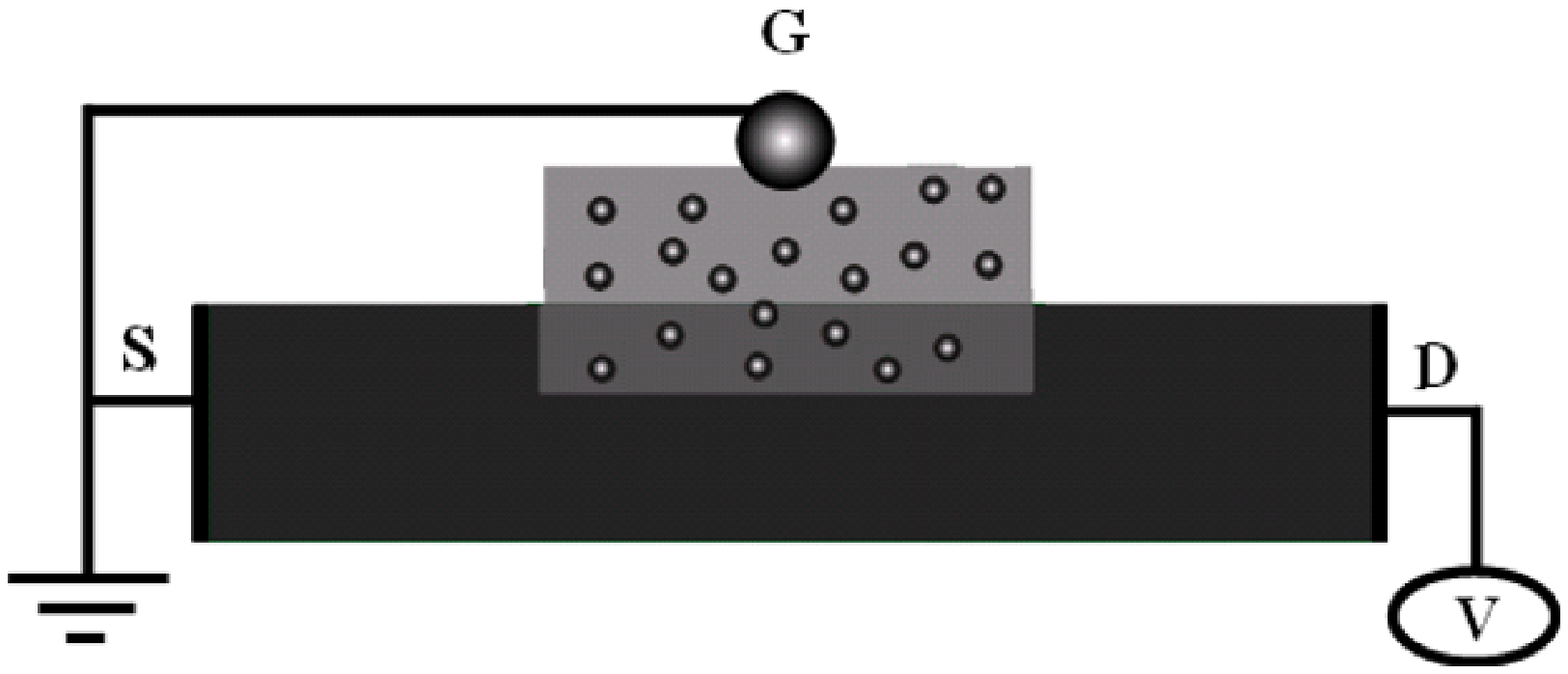,width=8cm,height=4cm}}
\subfloat[]{ \psfig{file=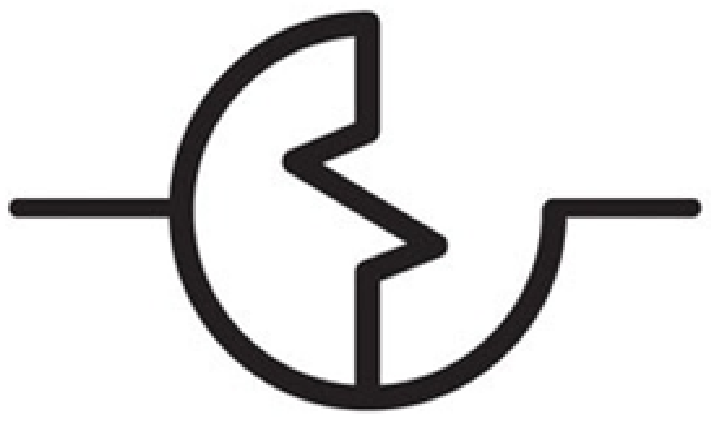,width=4cm,height=3cm}}\\

\end{center}
\caption[]{Schematic representation (a) and electronic symbol (b) of the organic memristive device.}
\label{Fig1}
\end{figure*}

The organic memristive element is shown schematically in Fig.~\ref{Fig1}(a). Its active zone is a heterojunction between the conducting polymer (polyaniline (PANI)) with the solid electrolyte (polyethylene oxide doped with lithium salt (PEO)). The working principle is based on the drastic difference of the PANI conductivity in its oxidized and reduced forms~\cite{12}. Redox reactions take place according to the actual potential of the PANI in the active zone with respect to  the reference potential of the silver wire in the PEO that is maintained, together with one metal electrode (source), at the ground level. Such device configuration implies that it can be considered as a two-terminal element regarding its connection to the external circuit. More details on the single device fabrication procedure can be found in previous works~\cite{13}. The symbol of the organic memristive device is shown in Fig.~\ref{Fig1}(b). Its shape is different from that introduced by Chua --- this difference is due to the fact that our element is an unisotropic system and cannot be represented by the symmetric symbol.

Voltage sweep and drain current measurements were performed with a 236 source measure unit (Keithley), while other current measurements were performed with a 6514 system electrometer (Keithley). Both Keithley units were linked to a PC and operated via MATLAB scripts, which allowed full automation of the measuring procedure.

\section{Results and Discussion}

\begin{figure}[t]
\begin{center}
\psfig{file=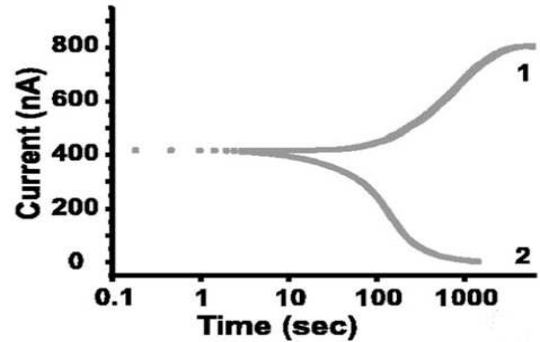, width=8cm,height=5cm} 
\end{center}
\caption{Experimental temporal behavior of the output current of the organic memristive device at constant bias voltage of +0.8 V (curve 1) and -0.3 V (curve 2).}
\label{Fig2}
\end{figure}

Before starting with logic circuits, it is necessary to recall a basic property of the organic memristor, allowing us to consider it as a synapse analogue. When biased positively (higher that oxidation potential), its conductivity increases over time until some saturation level. When biased negatively (reduction potential of PANI is approximately +0.1 V), it decreases its conductivity. In our first elements the conductivity ratio in saturated conducting and insulating states was about 100, while improvement in the material choice and device construction has resulted in the subsequent increase of this ratio to approximately 1000~\cite{14}. Experimental temporal dependences of the output current at fixed positive (higher than oxidation potential) and negative (any) voltages is shown in  Fig.~\ref{Fig2}. 

Differences in the kinetics is connected to the necessity to redistribute the potential during transformation of the conductivity profile in the active zone. The details of the model, explaining such behavior, can be found in~\cite{15}.  Briefly, the model is based on dividing the active zone into narrow strips, supposing that all processes occur simultaneously within a single strip.  A timer was attributed to each strip; at every time step, the model calculates the potential profile on the active zone. Strips that are between the oxidizing and reduction potentials maintain their current conductivity.  Strips with potentials outside of this constraint begin to vary their conductivity according to an exponential law --- constants were obtained by fitting experimental data.  The starting time of these processes is connected to the first moment when the strip attains an oxidizing/reduction potential.  This model was successfully applied for (i) explanation of voltage-current characteristics and (ii) kinetics of the conductivity variation. Moreover, it describes qualitively the behavior of the organic memristive device in current auto-oscillation mode~\cite{vic-oscillator}.  Later, the model was simplified to allow for application to more complex circuits composed of myriad memristive devices~\cite{pincella}. In this case, the memristor is represented as a variable resistor and a capacitor, representing the contact of conducting polymer with solid electrolyte.  Experimental data on the kinetics can be best fitted with two exponential functions~\cite{16}, expressed in (\ref{Eq1}).  Here, $T_{1}$ and $T_{2}$ are the time constants and C is a constant, corresponding to the conductivity value in the saturating state. This tendency is valid for both increase and decrease of the conductivity, but the values of the $A_{1}$ and $A_{2}$ constants can be different.

\begin{equation}
\displaystyle I=A_{1}e^{-\dfrac{t}{T_{1}}} + A_{2}e^{-\dfrac{t}{T_{2}}} + C
\label{Eq1}
\end{equation}

The reason of the double exponential function utilization is mainly due to the processes at the boundary between solid electrolyte and conducting polymer. One time constant could be just the RC of the junction, while the second one (slower) is connected to a drift of lithium ions from PEO to PANI and vice versa.  Three basic logic elements were fabricated and studied within this work.

\subsection{Memorized OR (MOR)}

\begin{figure}[t]
\begin{center}
\psfig{file=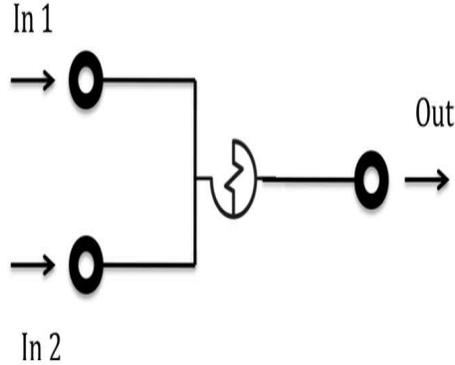, width=6cm,height=5cm} 
\end{center}
\caption{Scheme of the memorized OR element (MOR), realized with organic memristive device.}
\label{Fig3}
\end{figure}

This is the simplest logic element that can be realized as an organic memristive system. The configuration of the circuit requires the utilization of a single organic memristor and is shown in Fig.~\ref{Fig3}.

\begin{figure}[t]
\begin{center}
\subfloat[]{\psfig{file=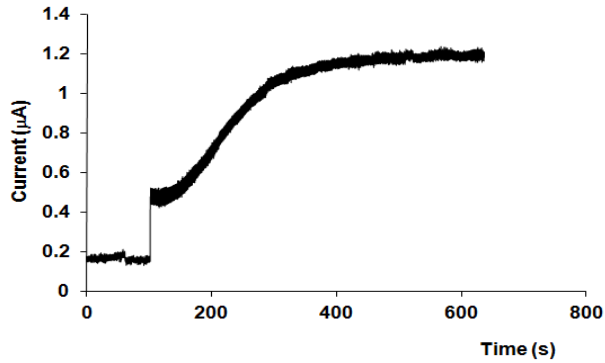, width=8cm,height=5cm}}\\
\subfloat[]{\psfig{file=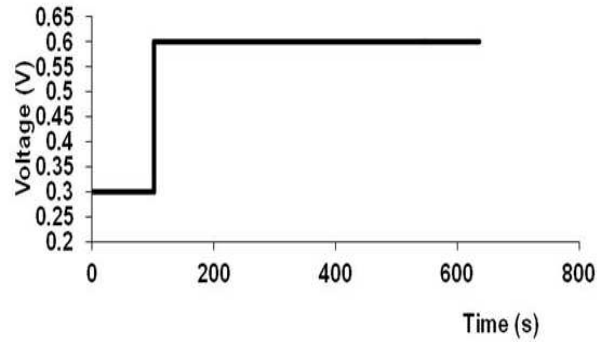,width=8cm,height=5cm}}
\end{center}
\caption{Temporal variation of the output signal of MOR element (a), variation of the voltage on one (any) input is shown in (b).}
\label{Fig4}
\end{figure}

Two independent inputs from separate voltage sources are connected to the memristor $D$ electrode (voltages) and output is the value of the current at $S$ electrode chain. The value of each input voltage value is sufficient to transfer the memristor to the conducting state. Therefore, the application of the any input signal over the oxidation potential will transfer the element to a more conducting state. Moreover, its value will depend on the duration of the applicaton of the inputs. Experimental dependence of the output variation as a function of any applied inputs is shown in  Fig.~\ref{Fig4}.

Thus, the function of logical OR is performed with the difference that the output signal ($S$) is restricted to the binary case, but can have any intermediate value according to the presence and duration of the input signal (see \ref{Eq2}), where $t_{1}$ and $t_{2}$ are total time intervals where each input was respectively activated, and $I_{out}(\infty)$ is the total possible current based on the current timestep.

\begin{equation}
\displaystyle S_{out}(t)=\dfrac{\displaystyle I_{out}(t_{1}+t_{2})}{\displaystyle I_{out}(\infty)}
\label{Eq2}
\end{equation}

Thus, the conductivity of this element and, therefore, the state at the output electrode will vary from 0 to 1 when at least one of the input is activated, and this state will be preserved until the next action (reinforcing or inhibiting) will be done.  Figs~\ref{Fig4},~\ref{Fig6} and~\ref{Fig8} present experimental data reporting output current values. In order to have $S_{out}(t)$ ($Y$ axis) varying from 0 to 1, they must be normalized to $I_{out}(\infty)$.

\subsection{Memorized AND (MAND)}

\begin{figure}[t]
\begin{center}
\psfig{file=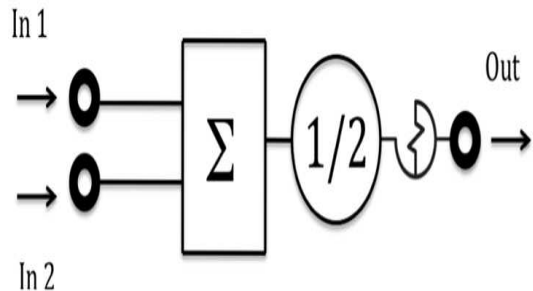, width=7cm,height=4cm} 
\end{center}
\caption{Circuit for the realization of MAND function.}
\label{Fig5}
\end{figure}

This function has a slightly more complicated realization than MOR as it demands to use additional elements. The scheme of the MAND circuit is shown in  Fig.~\ref{Fig5}.

\begin{figure}[t]
\begin{center}
\subfloat[]{\psfig{file=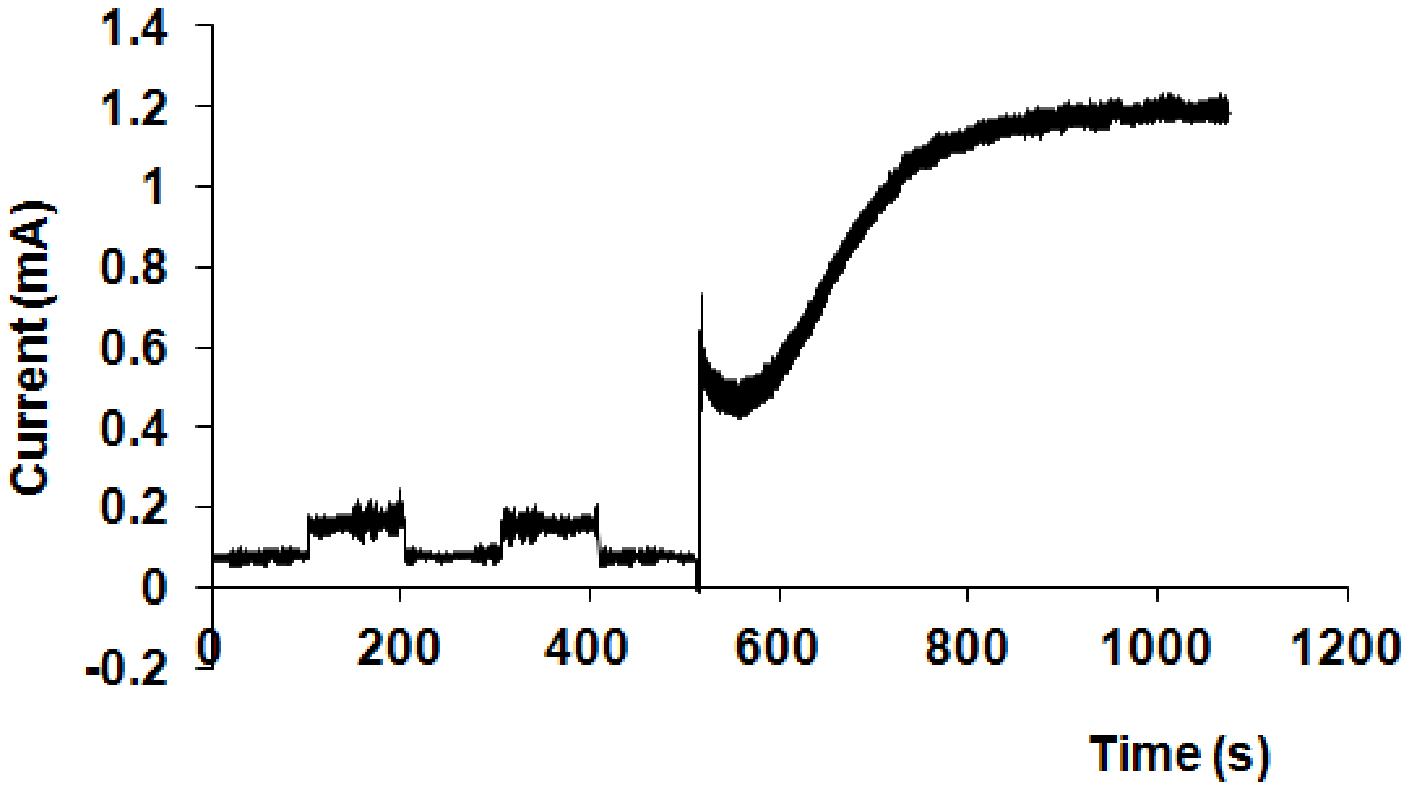, width=8cm,height=6cm}}\\
\subfloat[]{\psfig{file=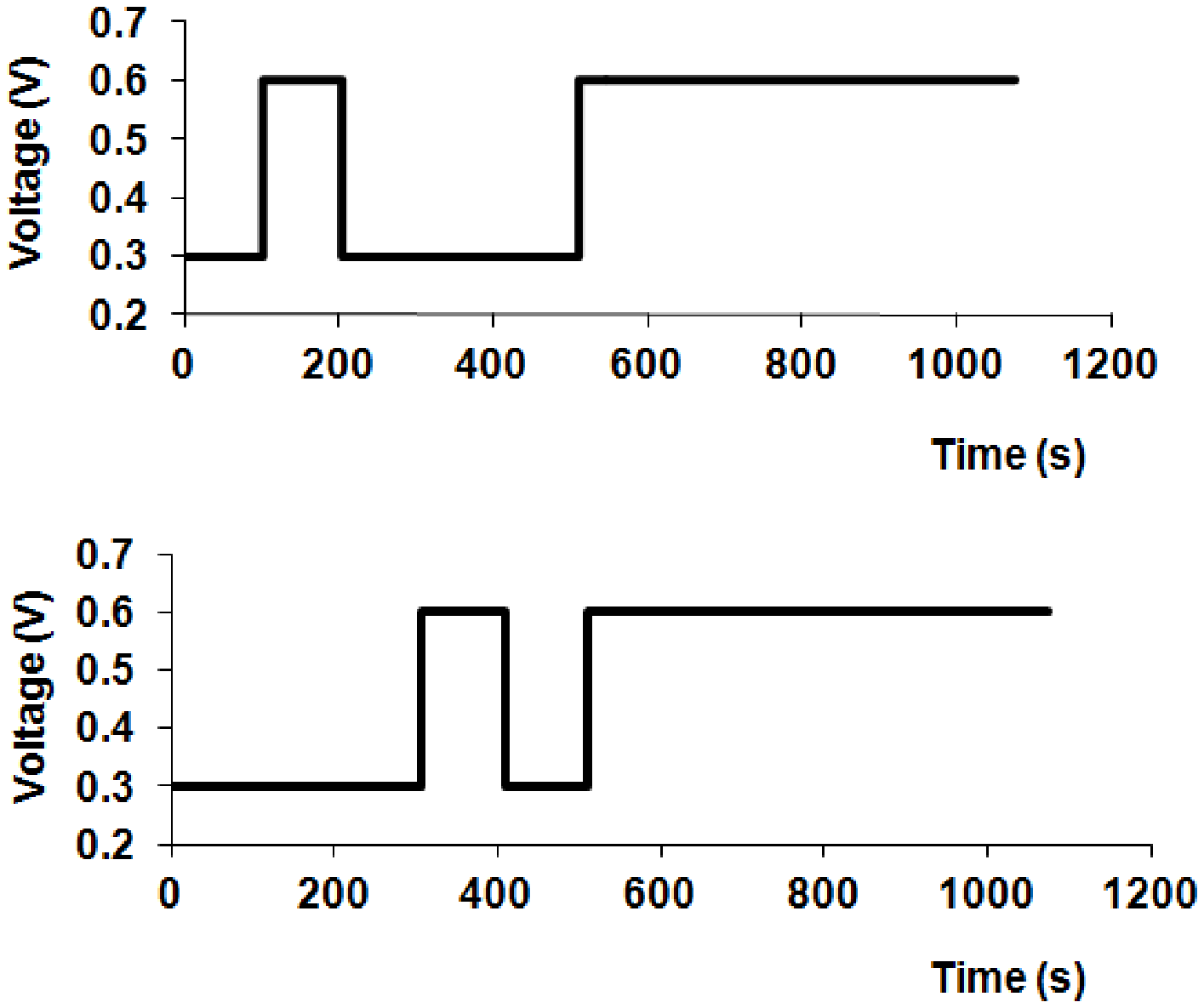,width=8cm,height=6cm}}
\end{center}
\caption{Temporal dependence of the output current of MAND element (a): dependences of the applied voltages to first and second inputs are shown in (b) (top) and (bottom) respectively.}
\label{Fig6}
\end{figure}

In the circuit we have two inputs connected to a summator (an operational amplifier where input voltages pass through equally-valued resistors), which combines the input voltages into a single voltage. In order to work with the same voltages as in the case of MOR element (each one is enough to transfer the memristor into the conducting state), we insert a dividing element after the summator which halves the summated input signal.  This element comprises two equally-valued resistors connected in serial, where the output is taken from a point between the two.  To decouple the dividing element from the input/output parts of the circuit, the resistors may be connected through operational amplifiers.  Experimental data, showing the status of the output current when each or both inputs are activated, are shown in  Fig.~\ref{Fig6}.  When both inputs are activated, we see a gradual increase of the output signal, while it remains constant when only one of them is activated. This element performs a function similar to synaptic associative learning. Reinforcement of the circuit demands the presence of both input signals.  When reinforcement occurs,  the conductivity of the circuit in the case of each individual input is increased, but the application of a single input cannot reinforce the conductivity of the circuit. The status of the output signal in this case can be described in (\ref{Eq3}), where $t_{comm}$ is the summarized time of all intervals when both input signals were activated simultaneously.\\

\begin{equation}
\displaystyle S_{out}(t)=\dfrac{\displaystyle I_{out}(t_{comm})}{\displaystyle I_{out}(\infty)}
\label{Eq3}
\end{equation}

\subsection{Memorized NOT (MNOT).}
	
\begin{figure}[t]
\begin{center}
\psfig{file=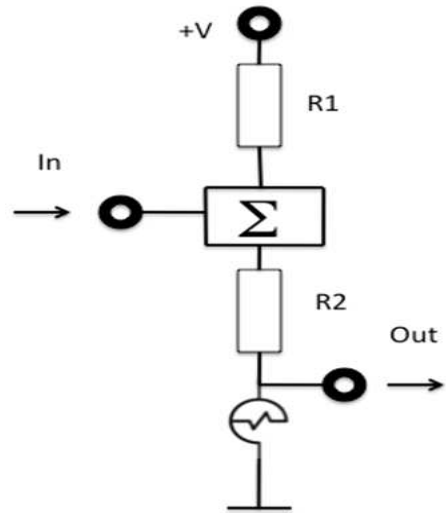, width=6cm,height=7cm} 
\end{center}
\caption{Scheme of the MNOT element based on the organic memristive device.}
\label{Fig7}
\end{figure}

The circuit capable of executing the NOT operation is shown in Fig.~\ref{Fig7}.  It contains two additional resistors, a summator and an external potential source (other than the source of the input signal). The value of the resistor $R_2$ must be an intermediate resistance between those of the memristor in the conducting and insulating state.  As the best organic memristive devices these walues can be 100 K$\Omega$ and 1000 M$\Omega$ respectively, $R_2$ can be equal to 10 M$\Omega$. The value of $R_1$ is not as critical, but it must be less than $R_2$. The external voltage $V_{con}$ must be small enough to prevent spontaneous transition of the memristor into the conducting state. 

\begin{figure}[t]
\begin{center}
\subfloat[]{\psfig{file=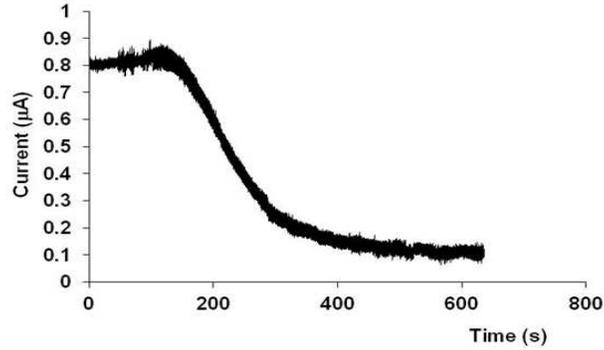, width=8cm,height=5cm}}\\
\subfloat[]{\psfig{file=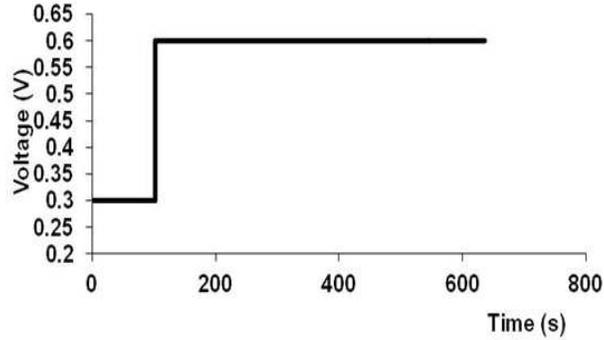,width=8cm,height=5cm}}
\end{center}
\caption{Temporal variation of the output signal of MNOT element (a), variation of the voltage on the single input is shown in (b).}
\label{Fig8}
\end{figure}

Before activating the input signal, practically all of the applied external potential will be distributed on the memristor as its resistance is two orders of magnitude higher than $R_1$ and $R_2$, providing high current at the output point through the low-valued external resistor.  Application of an input voltage of similar value to the previous gates will result in the transformation of the memristive device into the conducting state. Subsequent disabling of the input signal will redistribute the external potential through $R_{1}$, $R_{2}$ and the memristor, giving an output value as in (\ref{Eq3}), where $R_{M}(t)$ in  is the actual value of the memristors resistance.

\begin{equation}
\displaystyle S_{out}(t)=\dfrac{\displaystyle R_{M}(t)}{\displaystyle \left( R_{1} + R_{2} + R_{M}(t) \right)}
\label{Eq4}
\end{equation}

For the resistor values mentioned above it is possible to acheive a ratio of two orders of magnitude between the output signal before and after application of the input.  Temporal behavior observed in the experimentally realized circuit is shown in Fig.~\ref{Fig8}.  In addition to acting as a memorized NOT element, this circuit can be also considered as an inhibiting synapse analog as (i) application of the input signal results in depression of the output signal and (ii) the degree of the suppression depends on the duration of the input signal.

\section{Simulation: Constructing an Adder}

\begin{figure}[h!]
\begin{center}
\psfig{file=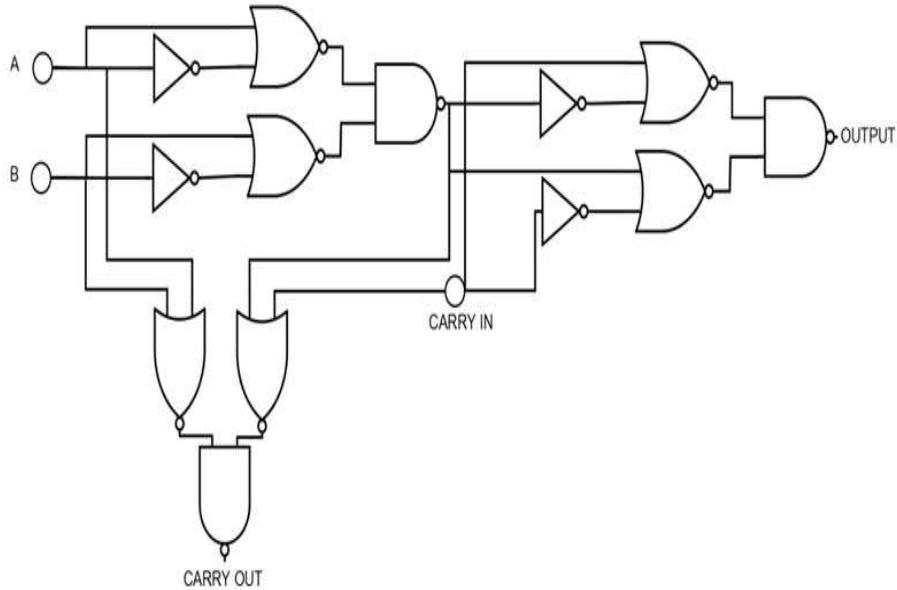, width=12cm,height=8cm} 
\end{center}
\caption{Gate-level schematic of the one-bit full adder consisting of MAND MOR and MNOT gates.  Traditional node symbols are used to represent their memristive counterparts.  Nodes A, B, and CARRY IN are input (logic 1 = 0.6V, logic 0 = 0.1V). OUTPUT and CARRY OUT nodes are also labelled.}
\label{Fig9}
\end{figure}

To provide a practical demonstration of a circuit comprised of memristive logic elements, we simulate the 1-bit full adder shown in  Fig.~\ref{Fig9}.  Adders are both widely used and functionally well-understood.  In addition, multiple adders of this type can be cascaded together to form more complex arithmetic units, increasing potential practical applications. Simulation parameters are based on recordings of physcially-realized gates.  For an overview of memristor-based arithmetic, \\see~\cite{mem-arithmetic}.

Inputs A and B, plus a CARRY IN (e.g. from the CARRY OUT of previous cascaded adders) can be set to logical 0 (0.1V) or logical 1 (0.6V).  The circuit is initialised by applying logical 0 to all inputs for the first 100ms, after which the chosen input pattern is applied for a further 300ms with a simulation time step of 1ms.  In the following simulation, we supply two distinct arbitrary input patterns to the circuit.  The first pattern is A = 0, B = 1, CARRY IN = 0, with the second being A=1, B = 0, CARRY IN = 1.  

Parameters are oxidation potential = 0.5V, reduction potential = -0.1V, $T_1$ = 30, $T_2$ = 300, $A_1$ = $-3\times10^{-7}$, $A_2$ = $-1\times10^{-7}$, C = $4\times10^{-7}$.  For MNOT gates, $R_1$ = $1\times10^{6}$$\Omega$, $R_2$ = $1\times10^{7}$$\Omega$, constant external voltage = 0.3V.  

As outputs tend to be measured as current, and inputs are always taken as voltage, we convert between the two using a proportionality constant B = $1.5\times10^{6}$, where V = IB.  In other words, the maximum attainable output current corresponds to a subsequent input voltage of 0.6V - in reality, insertion of appropriate resistors can provide this functionality.

\begin{figure*}[h]
\begin{center}
\subfloat[]{ \psfig{file=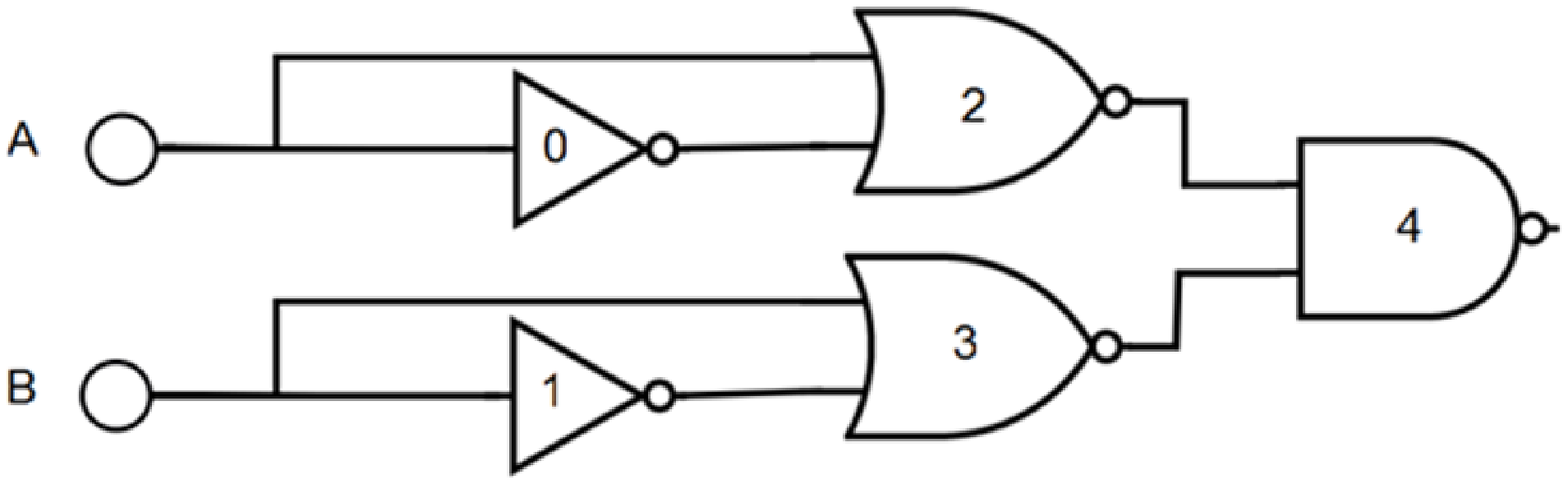,width=5.5cm,height=5.5cm}}\\
\subfloat[]{ \psfig{file=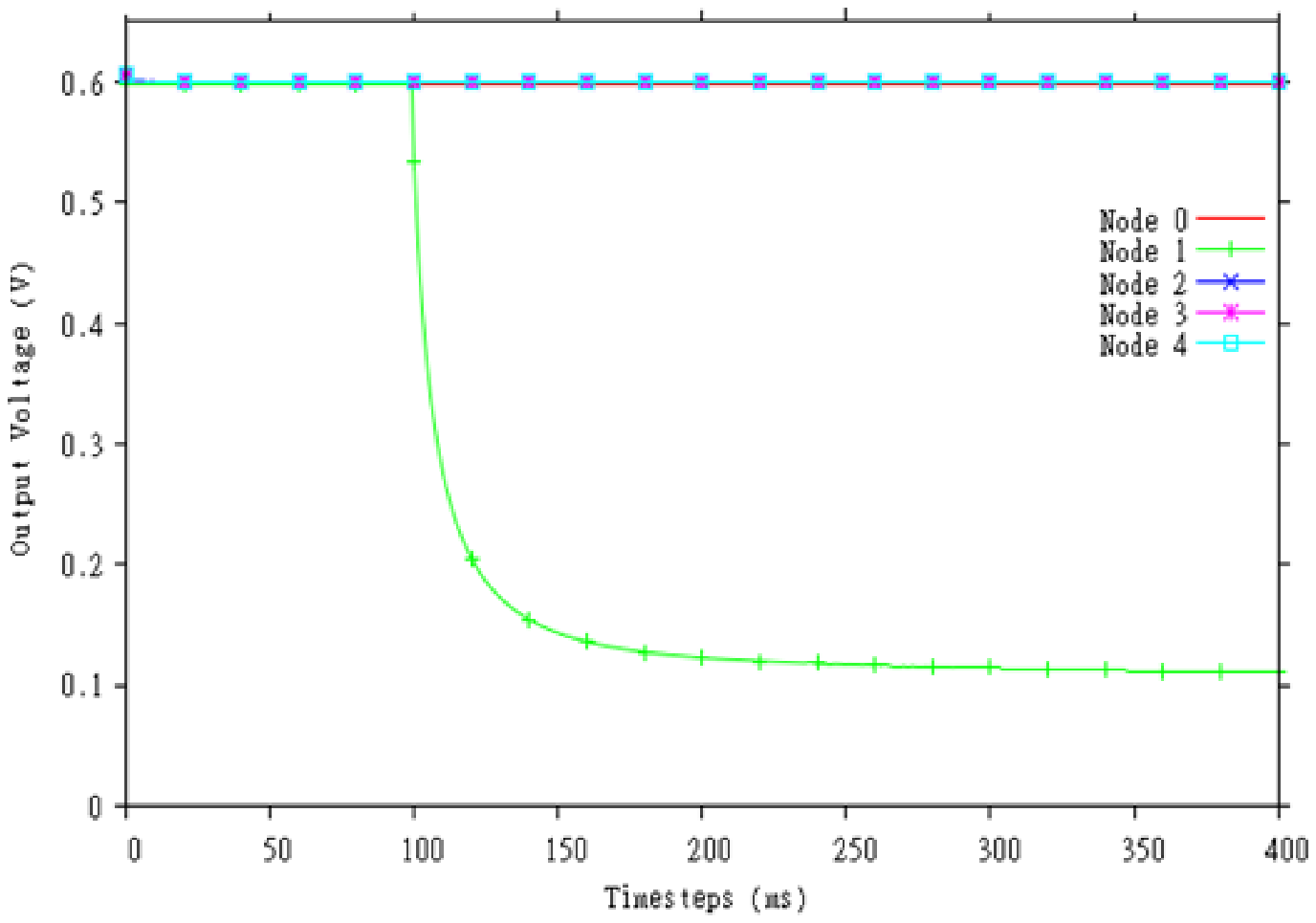,width=5.5cm,height=5.5cm}} 
\subfloat[]{ \psfig{file=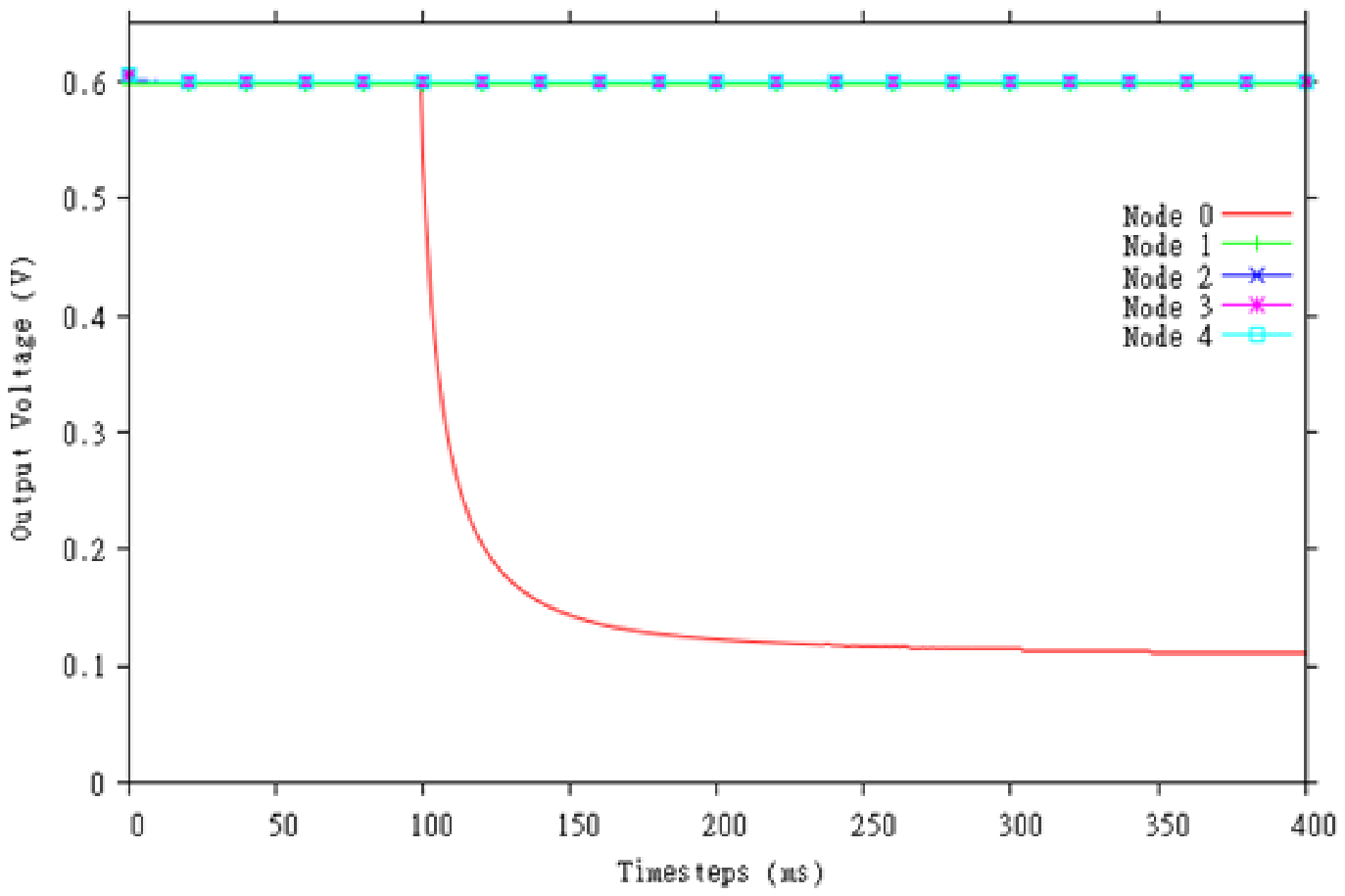,width=5.5cm,height=5.5cm}} \\
\end{center}
\caption[]{ (a) Half-adder 1 schematic (b) node output voltage response for input of 010 (c) node output voltage response for input for 101.  Node numbers allow for cross-referencing.}
\label{Fig10}
\end{figure*}

\begin{figure*}[h]
\begin{center}
\subfloat[]{ \psfig{file=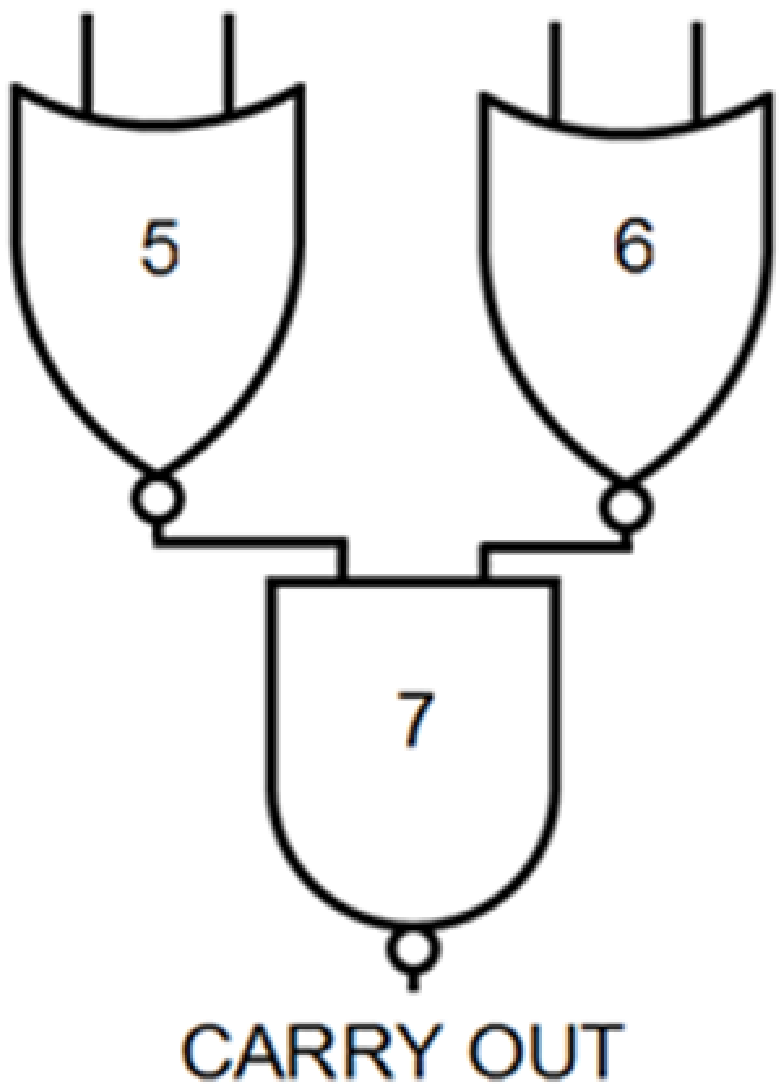,width=5cm,height=5cm}}\\
\subfloat[]{ \psfig{file=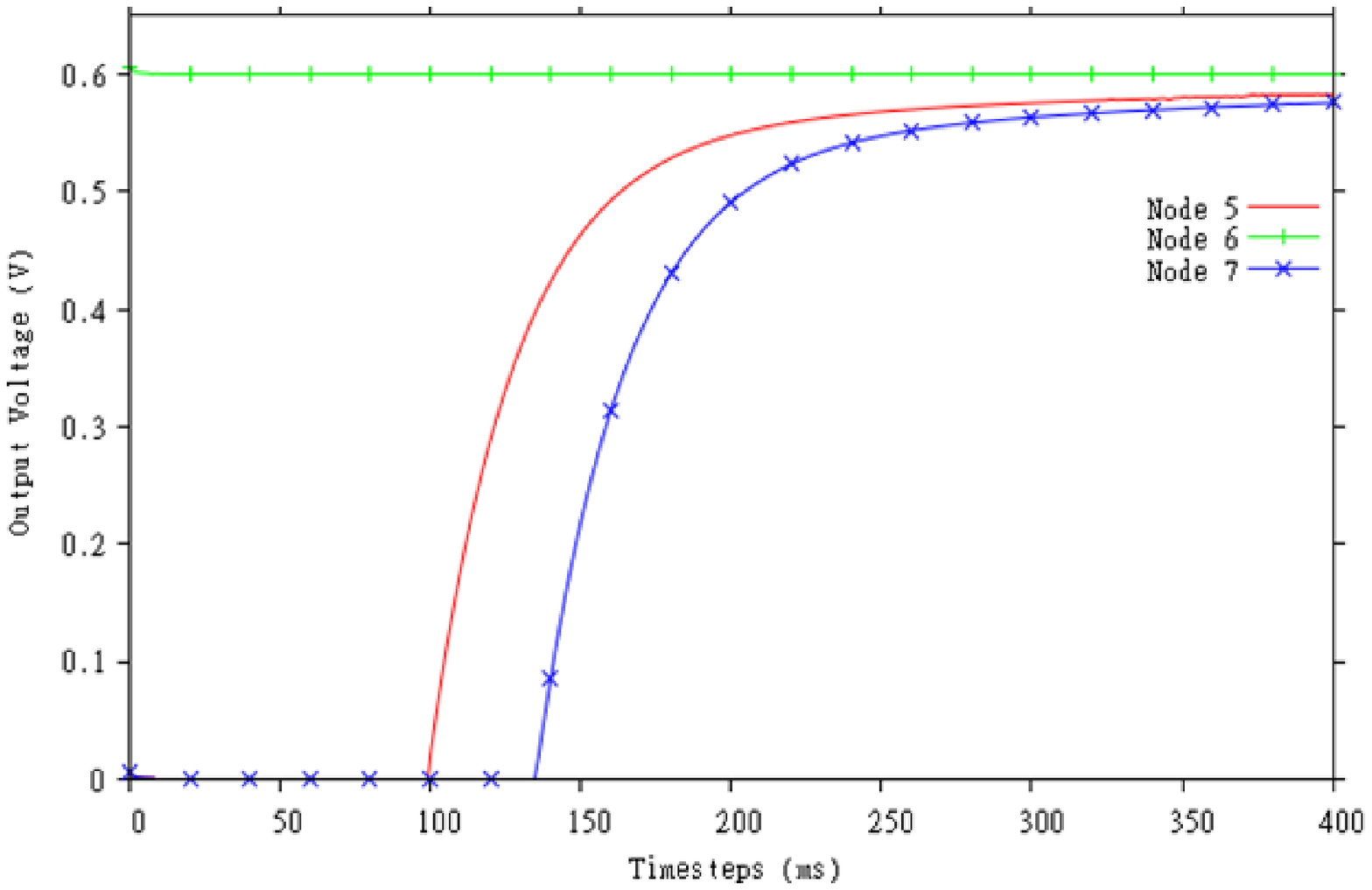,width=6cm,height=6cm}}
\subfloat[]{ \psfig{file=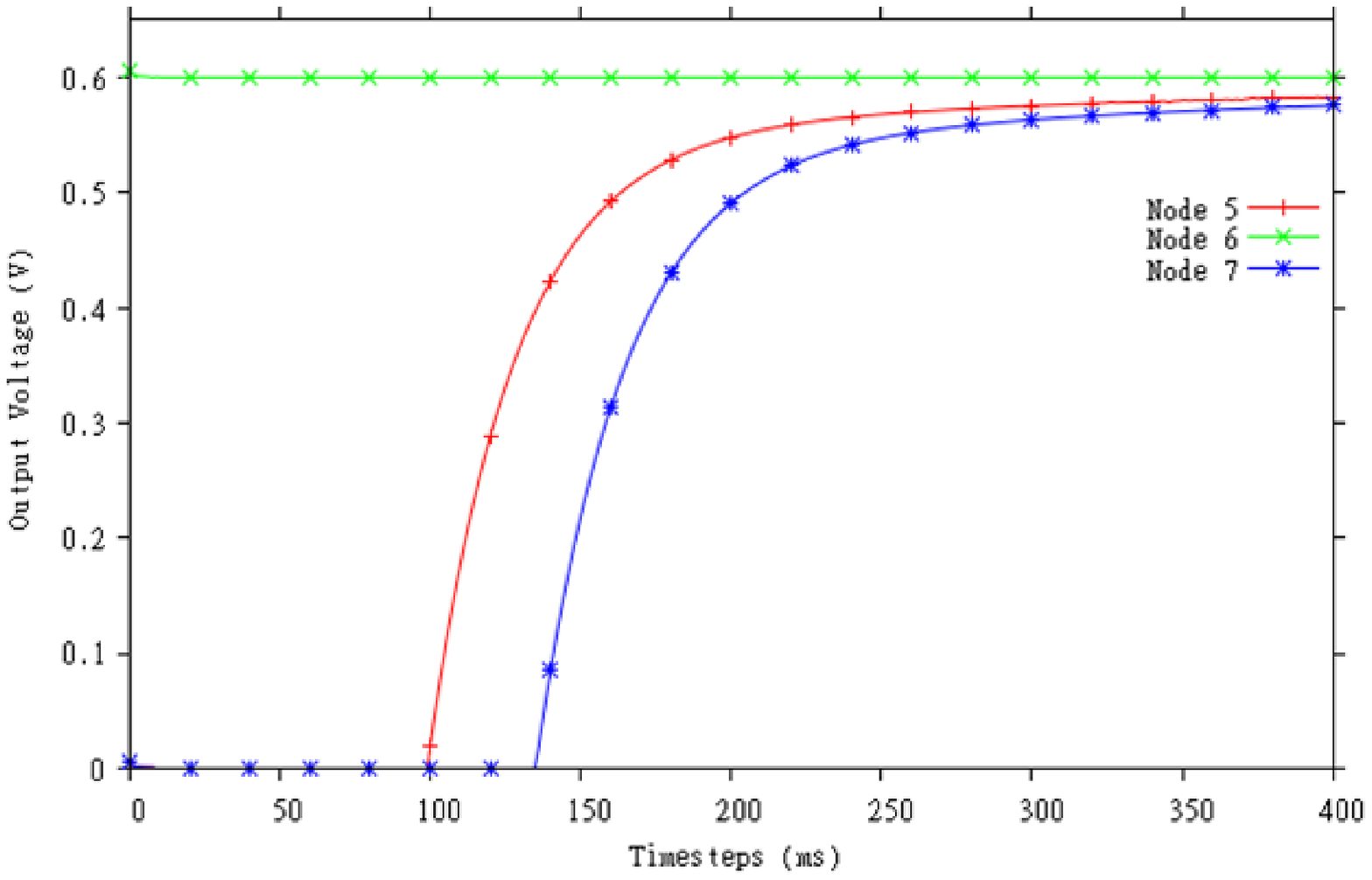,width=6cm,height=6cm}}
\end{center}
\caption[]{ (a) Carry unit schematic (b) node output voltage response for input of 010 (c) node output voltage response for input for 101.  Node numbers allow for cross-referencing.  Node 7 is the circuit carry out.}
\label{Fig11}
\end{figure*}

\begin{figure*}[h]
\begin{center}
\subfloat[]{ \psfig{file=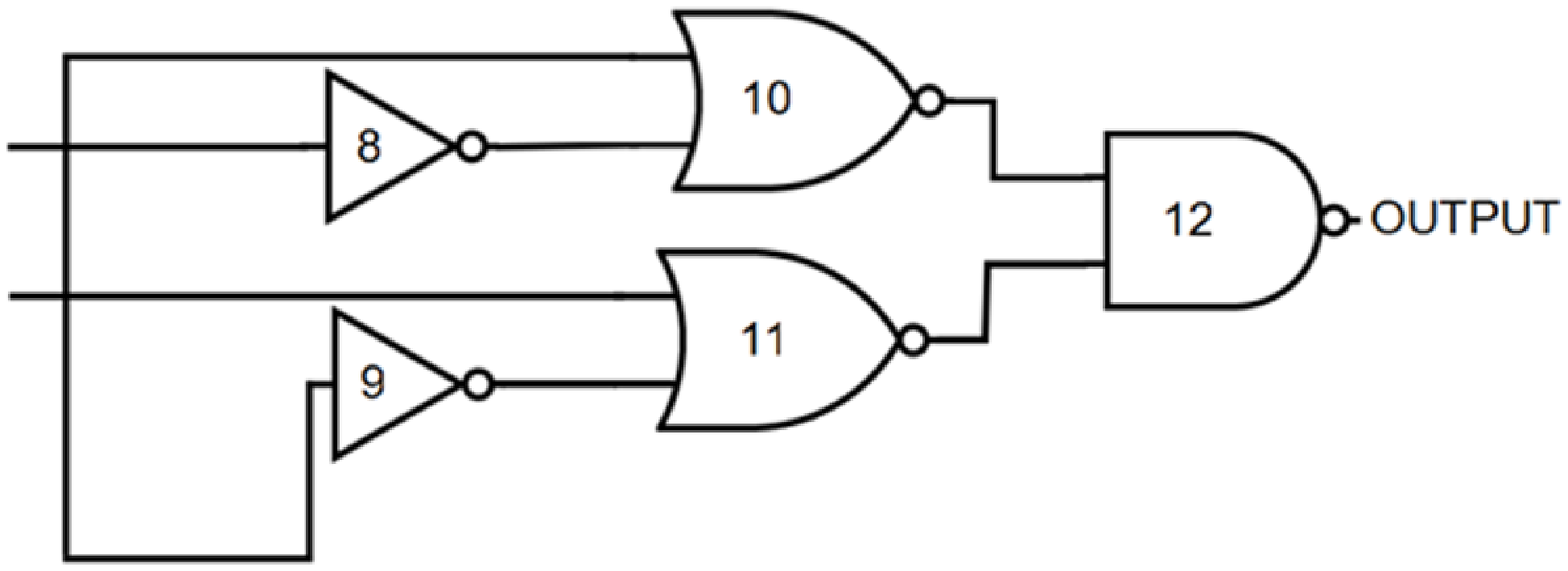,width=6cm,height=6cm}}\\
\subfloat[]{ \psfig{file=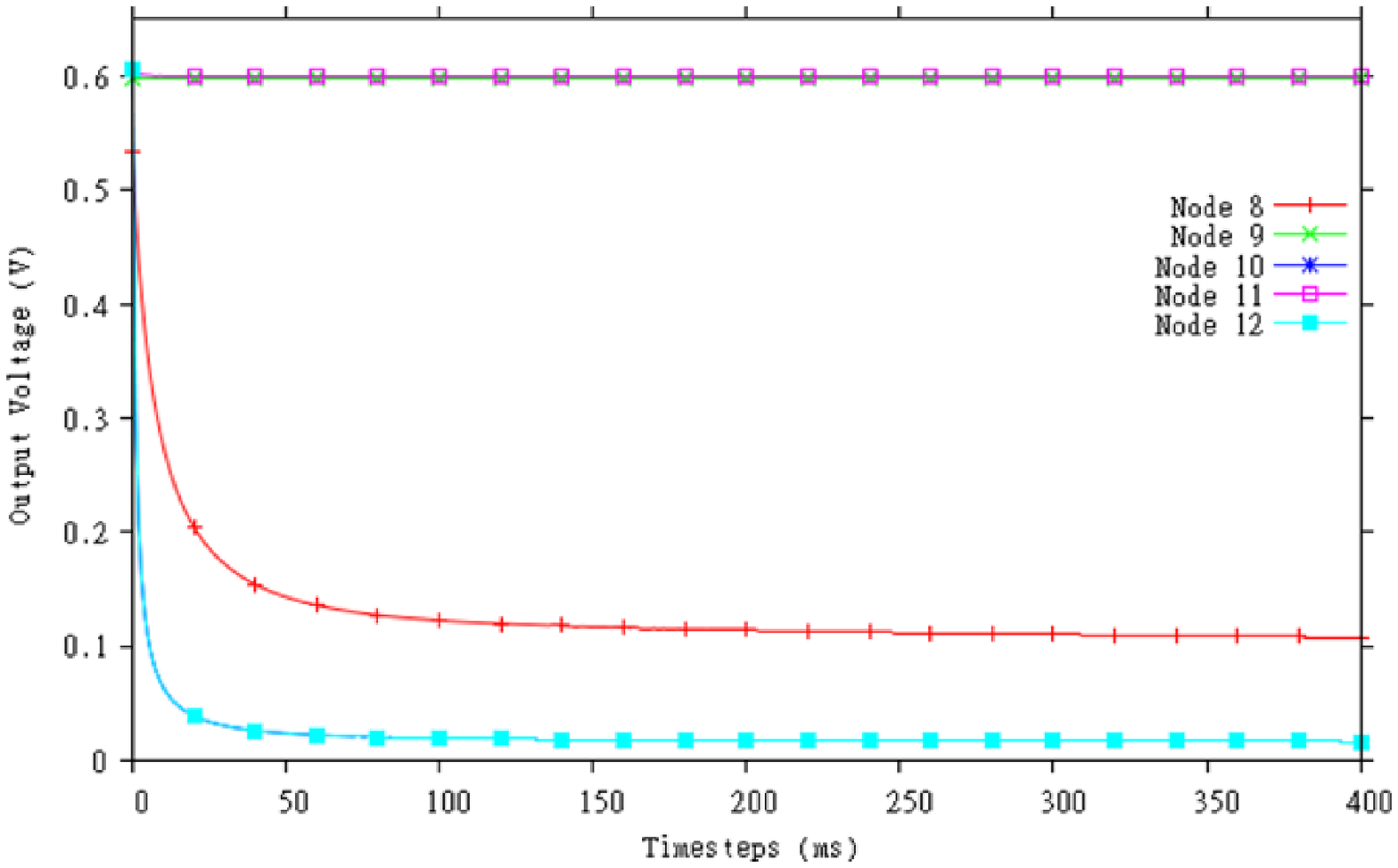,width=6cm,height=6cm}}
\subfloat[]{ \psfig{file=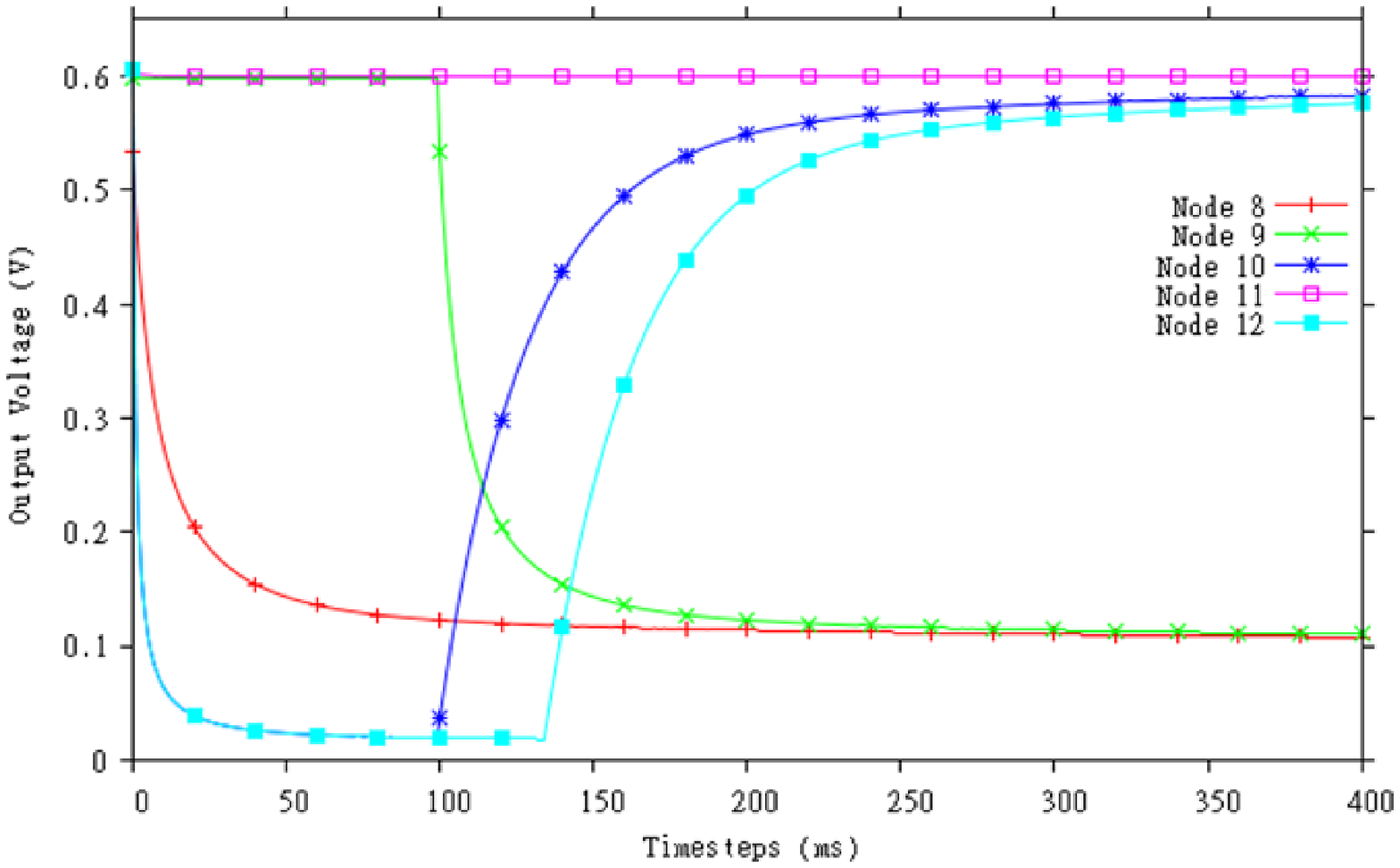,width=6cm,height=6cm}}
\end{center}
\caption[]{(a) Half-adder 2 schematic (b) node output voltage response for input of 010 (c) node output voltage response for input for 101.  Node numbers allow for cross-referencing.  Node 12 is the circuit output.}
\label{Fig12}
\end{figure*}

For the sake of clarity, we split the circuit into the two half-adders (Figs.~\ref{Fig10}(a) and~\ref{Fig12}(a)) and the carry calculator (Fig.~\ref{Fig11}(a)).  Figs.~\ref{Fig10}(b)~\ref{Fig11}(b) and~\ref{Fig12}(b) show output voltages for the respective subcircuit nodes for the first input sequence, Figs.~\ref{Fig10}(c)~\ref{Fig11}(c) and~\ref{Fig12}(c) show the same for the second input sequence.  All graphs show the first 400ms of simulation (100ms initialisation, 300ms input application).

We observe that after the first state input at 100ms, the output (node 12) takes a maximum of 100ms to reach an binary-thresholdable (e.g. $>$0.5V) output value (Fig.~\ref{Fig12}(c)).  Delays are mainly attributed to the requirement of previous nodes to reach an input voltage surpassing the oxidation potential --- especially MAND nodes as both inputs must surpass this threshold to alter the state of the node.

Pleasingly, no output voltages are ambiguous ($\approx$0.3V), indicating that successful binary operation will be possible given a sensible threshold level for reading logical 0/1 at the output node.  

MNOT gates never reach 0V, typically reaching a minimum of 0.104V --- low enough to pass a sensible threshold.  We note that this value decreases as the constant input voltage increases, however increasing the voltage too much could lead to the node changing in the absence of an appropriate input signal in a real (e.g. noisy) environment.  Increasing the constant input voltage brings the additional advantage of reducing the response time of the node when presented with a steadily increasing input voltage.

\section{Conclusions}

In this work we have suggested and fabricated several logic elements with memory. Their characteristic feature is that when inputs are activated, they behave as analog logic elements whose output signal can vary gradually from is 0 to 1  according to the duration of the applied input signals.  We have also suggested an architecture of a cascadable adder based on these realized elements and simulated its properties.

It is possible in principle to realize a system where these logic elements will return to their initial states after predefined time intervals, by applying periodic inverted (negative) potentials to all inputs.  Such action will bring all the memristors back to the insulating state and will make the realized circuits more similar to traditional electronic logic elements. 

Given the property of memristance, it seems more interesting to design new systems using these new functional properties e.g. the capability to memorize the history of inputs to the device.  In this respect, the output will be not a simple binary decision, but some intermediate value, that depends on the duration of the inputs. Such processing is to some degree similar to a brainlike logical decision making process: an answer (YES / NO) depends not only on the configuration of external stimuli, but also on past experience.  In addition, the realized elements can be useful for non-linear dynamic systems. Being assembled in multielement circuits, the response will strongly depend on the history of the system evolution due to the cross-talk between elements~\cite{13}. Moreover, simple modification of the memristor device with a charge-accumulating element will transform the device into an oscillator~\cite{vic-oscillator}; in this case dynamically ocsillating networks may be realized.

\section*{Acknowledgments} \noindent This work was supported by EPSRC Grant no. EP/H014381/1.

\end{document}